\documentclass[preprint,pra,onecolumn]{revtex4}%
\usepackage{amsfonts}
\usepackage{amsmath}
\usepackage{amssymb}
\usepackage{graphicx}%
\begin{document}
\title{Entanglement changing power of two-qubit unitary operations}
\author{Ming-Yong Ye}
\email{myye@mail.ustc.edu.cn}
\author{Dong Sun}
\author{Yong-Sheng Zhang}
\email{yshzhang@ustc.edu.cn}
\author{Guang-Can Guo}
\email{gcguo@ustc.edu.cn}
\affiliation{Key Laboratory of Quantum Information, Department of Physics, University of
Science and Technology of China (CAS), Hefei 230026, People's Republic of China}

\begin{abstract}
We consider a two-qubit unitary operation along with arbitrary local unitary
operations acts on a two-qubit pure state, whose entanglement is $C_{0}$. We
give the conditions that the final state can be maximally entangled and be
non-entangled. When the final state can not be maximally entangled, we give
the maximal entanglement $C_{\max}$ it can reach. When the final state can not
be non-entangled, we give the minimal entanglement $C_{\min}$ it can reach. We
think $C_{\max}$ and $C_{\min}$ represent the entanglement changing power of
two-qubit unitary operations. According to this power we define an order of gates.

PACS number(s): 03.67.-a, 03.67.Mn, 03.65.Ta

\end{abstract}
\maketitle

\section{introduction}

Entanglement is a fundamental resource in quantum information theory. It is
used in quantum key distribution \cite{ekert}, dense coding \cite{bw92},
teleportation \cite{bbc} and so on. Since entanglement is such a valuable
resource, many efforts are devoted to generate it \cite{sackett,pan} and
quantify it \cite{yu}. Recently some researchers begin to investigate nonlocal
operations \cite{dvc,dvclp,vc,nielsenetal}. On the one hand, nonlocal
operations can generate entanglement. On the other hand, entanglement can be
used to implement nonlocal operations if local operations and classical
communication are permitted \cite{ejpp,rag,cdkl}.

Entanglement can be generated by nonlocal operations, so we should ask
entangling capacity of nonlocal operations, nonlocal Hamiltonians or unitary
gates. Some results have been derived\cite{dvclp,ws,zzf}, especially Kraus and
Cirac \cite{kc} calculate the maximal final entanglement after a two-qubit
gate along with arbitrary local unitary operations acted on an initial
non-entangled state. Leifer \textit{et al. }\cite{lhl} consider a similar
question. Their efforts are devoted to maximize the entanglement of the final
state minus the entanglement of the initial state. They think this quantity
represents the entanglement generating ability of a nonlocal gate. In this
paper we consider a general question. Suppose the entanglement of the initial
pure state is given, denoted by $C_{0}$, we want to know the reachable maximal
entanglement $C_{\max}$ of the final state after a two-qubit gate acted on the
initial state, where local unitary operations can be freely used. Obviously
Kraus and Cirac \cite{kc} solved the question where $C_{0}$ is zero. We solve
the question for a general $C_{0}$. In this paper we also calculate the the
minimal reachable entanglement $C_{\min}$ of the final state for a general
$C_{0}$. The minimal entanglement can be zero if measurements are permitted,
but we still consider it mathematical interest. We think $C_{\max}$ and
$C_{\min}$ represent the entanglement changing power of a two-qubit gate. The
entanglement of the final state can be any value between them due to continuity.

The structure of the paper is as follows. In Sec. II we introduce concurrence
\cite{wootters} and canonical decomposition of two-qubit gates \cite{kc}. We
use concurrence to quantify two-qubit entanglement and use canonical
decomposition to classify two-qubit gates. In Sec. III we calculate the
minimal final entanglement after a two-qubit gate acted on a maximal entangled
state. This result will be used to judge whether the maximal final
entanglement $C_{\max}$ can be $1$ or not for a general $C_{0}$. In Sec. IV we
calculate the maximal and minimal entanglement of the final state for a
general initial state. Finally we conclude this paper in Sec. V and define an
order of gates.

\section{concurrence and canonical decomposition}

Concurrence \cite{wootters} is defined to quantify entanglement of formation
of mixed two-qubit states. For pure states it has a simple form. We write
two-qubit states in magic basis $\left\vert \Psi\right\rangle =\sum_{k=1}%
^{4}b_{k}\left\vert \Phi_{k}\right\rangle $, then the concurrence $C\left(
\left\vert \Psi\right\rangle \right)  =\left\vert \sum_{k=1}^{4}b_{k}%
^{2}\right\vert $, where $\left\{  \left\vert \Phi_{k}\right\rangle \right\}
_{k=1}^{4}$ is defined as follows,%
\begin{equation}
\left\vert \Phi_{1}\right\rangle =\frac{-i}{\sqrt{2}}\left(  \left\vert
00\right\rangle -\left\vert 11\right\rangle \right)  , \label{1}%
\end{equation}%
\begin{equation}
\left\vert \Phi_{2}\right\rangle =\frac{1}{\sqrt{2}}\left(  \left\vert
00\right\rangle +\left\vert 11\right\rangle \right)  , \label{2}%
\end{equation}%
\begin{equation}
\left\vert \Phi_{3}\right\rangle =\frac{-i}{\sqrt{2}}\left(  \left\vert
01\right\rangle +\left\vert 10\right\rangle \right)  , \label{3}%
\end{equation}%
\begin{equation}
\left\vert \Phi_{4}\right\rangle =\frac{1}{\sqrt{2}}\left(  \left\vert
01\right\rangle -\left\vert 10\right\rangle \right)  . \label{4}%
\end{equation}
The concurrence $C$ is zero iff the two-qubit state is a product state. When
the state is maximally entangled the concurrence is $1$, which requires the
coefficients $\left\{  b_{k}\right\}  _{k=1}^{4}$ are real, except for a
global phase.

Now we introduce canonical decomposition of two-qubit unitary operations
\cite{kc}. Any unitary operation acting on two qubits has $15$ parameters but
it can be locally equivalent to an operation which only has $3$ parameters.
According to the canonical decomposition given by Kraus and Cirac \cite{kc},
we can decompose $U_{AB}=\left(  U_{A}\otimes U_{B}\right)  U_{d}\left(
V_{A}\otimes V_{B}\right)  $, where $U_{A},U_{B},V_{A}$ and $V_{B}$ are local
unitary operations and $U_{d}$ has a special form%
\begin{equation}
U_{d}=\exp\left(  i\sum_{j=1}^{3}\alpha_{j}\sigma_{j}^{A}\otimes\sigma_{j}%
^{B}\right)  , \label{5}%
\end{equation}
where $\pi/4\geqslant\alpha_{1}\geqslant\alpha_{2}\geqslant\left\vert
\alpha_{3}\right\vert \geqslant0$ and $\sigma_{1,2,3}$ are Pauli matrix.
Because local unitary operations do not change the entanglement, we only
discuss the entanglement changing power of $U_{d}$ instead of $U_{AB}$ in the
following. In fact we can always take $\alpha_{3}\geqslant0$ when we discuss
entanglement changing power \cite{kc,lhl}. And because entanglement is
invariant under conjugation, the entanglement changing power of $U_{d}$ is the
same as $U_{d}^{\ast}\left(  U_{d}^{\dagger}\right)  $. This means if $U_{d}$
can change the states of entanglement $C_{1}$ to the states of entanglement
$C_{2}$, conversely it can change the states of entanglement $C_{2}$ to the
states of entanglement $C_{1}$. This result will be used in the following. A
very important character of $U_{d}$ is that the magic basis states are its
eigenstates, $U_{d}\left\vert \Phi_{j}\right\rangle =e^{i\lambda_{j}%
}\left\vert \Phi_{j}\right\rangle $, where
\begin{equation}
\lambda_{1}=-\alpha_{1}+\alpha_{2}+\alpha_{3}, \label{6}%
\end{equation}%
\begin{equation}
\lambda_{2}=+\alpha_{1}-\alpha_{2}+\alpha_{3}, \label{7}%
\end{equation}%
\begin{equation}
\lambda_{3}=+\alpha_{1}+\alpha_{2}-\alpha_{3}, \label{8}%
\end{equation}%
\begin{equation}
\lambda_{4}=-\alpha_{1}-\alpha_{2}-\alpha_{3}. \label{9}%
\end{equation}

\section{the case that the initial state is maximally entangled}

In this section we consider the situation where $C_{0}$ is $1$. We want to
know the maximal and minimal entanglement of the final state after a two-qubit
unitary gate acted on an initial maximally entangled state. Because the Bell
states are eigenstates of the unitary operation $U_{d}$, we can easily find
that the maximal entanglement of the final state $C_{\max}$ is $1$. To find
the minimal entanglement of the final state, we write the initial state in the
magic state, $\left\vert \Psi_{0}\right\rangle =\sum_{j=1}^{4}b_{j}\left\vert
\Phi_{j}\right\rangle $. The coefficients $\left\{  b_{j}\right\}  _{j=1}^{4}$
are real and $\sum_{j=1}^{4}b_{j}^{2}=1$. The final state is%
\begin{equation}
\left\vert \Psi\right\rangle =U_{d}\left\vert \Psi_{0}\right\rangle
=\sum_{j=1}^{4}b_{j}e^{i\lambda_{j}}\left\vert \Phi_{j}\right\rangle .
\label{10}%
\end{equation}
We want to minimize the concurrence $C$ of the final state $\left\vert
\Psi\right\rangle $. We define a Lagrangian function%
\begin{align}
L  &  =C^{2}-\mu\left(  \sum_{k=1}^{4}b_{k}^{2}-1\right) \label{11}\\
&  =\left(  \sum_{k=1}^{4}b_{k}^{2}e^{2i\lambda_{k}}\right)  \left(
\sum_{l=1}^{4}b_{l}^{2}e^{-2i\lambda_{l}}\right)  -\mu\left(  \sum_{k=1}%
^{4}b_{k}^{2}-1\right)  ,\nonumber
\end{align}
where $\mu$ is a Lagrangian multiplier, which is real. Differentiating gives%
\begin{equation}
\frac{\partial L}{\partial b_{j}}=2b_{j}e^{2i\lambda j}\left(  \sum_{l=1}%
^{4}b_{l}^{2}e^{-2i\lambda_{l}}\right)  +2b_{j}e^{-2i\lambda j}\left(
\sum_{k=1}^{4}b_{k}^{2}e^{2i\lambda_{k}}\right)  -2\mu b_{j}=0, \label{12}%
\end{equation}
multiplying $b_{j}$ and summing over $j$ gives%
\begin{equation}
\mu=2C^{2}. \label{13}%
\end{equation}
We write $\left(  \sum_{k=1}^{4}b_{k}^{2}e^{2i\lambda_{k}}\right)  =Ce^{i\eta
}$, then from Eq. (12) we get%
\begin{equation}
b_{j}C\cos\left(  2\lambda_{j}-\eta\right)  =b_{j}C^{2}. \label{14}%
\end{equation}
If $C$ is equal to $0$, this means $U_{d}$ can change the maximal entangled
states to the product states. Actually this question has been solved by Kraus
and Cirac \cite{kc} though they considered a different question. We write the
result here: if%
\begin{equation}
\alpha_{1}+\alpha_{2}\geqslant\pi/4\text{ and }\alpha_{2}+\alpha_{3}%
\leqslant\pi/4, \label{15}%
\end{equation}
then the two-qubit unitary operation $U_{d}$ can change maximally entangled
pure qubit states to product states along with local unitary operations. In
the following we focus on the cases where $C$ is not $0$. Now the Eq. (14)
becomes
\begin{equation}
b_{j}\cos\left(  2\lambda_{j}-\eta\right)  =b_{j}C. \label{16}%
\end{equation}
One possible solution of the Eq. (16) is $b_{j}=0$. But there must have some
nonzero coefficients. If there are only one nonzero coefficient, the initial
state is the eigenstate of $U_{d}$ and the final state is also a maximally
entanglement state. So there are at least two nonzero coefficients. Suppose
$b_{k}\neq0$ and $b_{l}\neq0$, then we have%
\begin{equation}
\cos\left(  2\lambda_{k}-\eta\right)  =\cos\left(  2\lambda_{l}-\eta\right)
=C. \label{17}%
\end{equation}
This means
\begin{equation}
\lambda_{k}-\lambda_{l}=n\pi\text{ or }\lambda_{k}+\lambda_{l}-\eta=n\pi,
\label{18}%
\end{equation}
where $n$ is an integer. Suppose no parameters $\lambda_{k}$ are equal. From
the value range of $\alpha_{1,2,3}$ we can find that $\lambda_{k}-\lambda
_{l}=n\pi$ is impossible. If there is another coefficient $b_{m}$ is also
nonzero, it will satisfy $\lambda_{k}+\lambda_{m}-\eta=n^{\prime}\pi$ for some
integer $n^{\prime}$. Then we can easily find that $\lambda_{l}-\lambda
_{m}=\left(  n-n^{\prime}\right)  \pi$, but this is impossible. So there are
only one pair of $\lambda_{k}^{\prime}s$ satisfy the Eq. (18). That is to say
there are only two coefficients which are nonzero and we denote them $b_{k}$
and $b_{l}$. Now our purpose is to minimize the concurrence $C=\left\vert
b_{k}^{2}e^{2i\lambda_{k}}+b_{l}^{2}e^{2i\lambda_{l}}\right\vert $ of the
final state under the condition $b_{k}^{2}+b_{l}^{2}=1$. Because%
\begin{align}
C^{2}  &  =b_{k}^{4}+b_{l}^{4}+2b_{k}^{2}b_{l}^{2}\cos\left(  2\lambda
_{k}-2\lambda_{l}\right) \label{19}\\
&  \geqslant\left\vert \cos\left(  \lambda_{k}-\lambda_{l}\right)  \right\vert
^{2},\nonumber
\end{align}
so the minimal entanglement of the final state is $C_{\min}=\min
_{k,l}\left\vert \cos\left(  \lambda_{k}-\lambda_{l}\right)  \right\vert $,
which is achieved when $b_{k}^{2}=b_{k}^{2}=1/2$. This minimal entanglement
$C_{\min}$ is calculated when we suppose no parameters $\lambda_{k}$ are
equal. Suppose the point $\left(  \alpha_{10},\alpha_{20},\alpha_{30}\right)
$ in parameter space makes some parameters $\lambda_{k}$ equal, but for
arbitrary small positive number $\xi$, there always exists some point $\left(
\alpha_{1}^{\prime},\alpha_{2}^{\prime},\alpha_{3}^{\prime}\right)  $ which
can not make any two parameters $\lambda_{k}$ equal, where $\left\vert
\alpha_{1}^{\prime}-\alpha_{10}\right\vert +\left\vert \alpha_{2}^{\prime
}-\alpha_{20}\right\vert +\left\vert \alpha_{3}^{\prime}-\alpha_{30}%
\right\vert <\xi$. So this constraint can be removed by continuity.

The result in this section can be applied in gate simulation. A maximal
entangled state can be used to implement deterministic controlled unitary
operations if local operations and classical communication (LOCC) are
permitted \cite{ejpp,rag}. If a two-qubit unitary gate can change some product
initial state into a maximal entangled state, then it can be used to simulate
controlled unitary operations under LOCC. If the non-local operation can not
change some product state into a maximal entangled one, we can let it act on
an initially entangled state to get a maximally entangled one. Then what is
the minimum entanglement of the initial state? It is $C_{\min}$ we calculate
in this section. We emphasize that if we can use ancillas the situation will
be different. For example, swap gate can not change a nonmaximally entangled
state into a maximal one without ancillas, but it can produce two maximally
entangled states from product states when ancillas are permitted.

\section{entanglement changing power}

Assume that the entanglement of the initial pure two-qubit state is $C_{0}$. A
nonlocal $U_{d}$ acts on this state and we want to know the possible maximal
and minimal entanglement of the final state. Some results have been derived
and we list them here:

When $C_{0}$ is $0$ \cite{kc}, the minimal entanglement of the final state
$C_{\min}$ is $0$ \cite{explain1}. The maximal entanglement of the final state
$C_{\max}$ is $1$ if $\alpha_{1}+\alpha_{2}\geqslant\pi/4$ and $\alpha
_{2}+\alpha_{3}\leqslant\pi/4$, otherwise it is $\max_{k,l}\left\vert
\sin\left(  \lambda_{k}-\lambda_{l}\right)  \right\vert $. This maximal
entanglement of the final state has special use and we name it $C_{0\max}$.

When $C_{0}$ is $1$, the maximal entanglement of the final state $C_{\max}$ is
$1$. The minimal entanglement of the final state $C_{\min}$ is $0$ if
$\alpha_{1}+\alpha_{2}\geqslant\pi/4$ and $\alpha_{2}+\alpha_{3}\leqslant
\pi/4$, otherwise it is $\min_{k,l}\left\vert \cos\left(  \lambda_{k}%
-\lambda_{l}\right)  \right\vert $. This minimal entanglement of the final
state also has special use and we name it $C_{1\min}$.

For a general $C_{0}$, we first want to know whether the entanglement of the
final state can be $1$ and $0$. Now the question is easy to answer. If
$C_{0}\geqslant C_{1\min}$, then the entanglement of final state can be $1$.
If $C_{0}\leqslant C_{0\max}$, then the entanglement of final state can be
$0$. So in the following we do not concern the question where the entanglement
of final state is $1$ or $0$.

We write the initial state in the magic basis: $\left\vert \Psi_{0}%
\right\rangle =\sum_{j=1}^{4}b_{j}\left\vert \Phi_{j}\right\rangle $. The
coefficients satisfy two conditions: $\sum_{j=1}^{4}\left\vert b_{j}%
\right\vert ^{2}=1$ and $\left\vert \sum_{j=1}^{4}b_{j}^{2}\right\vert
^{2}=C_{0}^{2}$. We want to calculate the possible maximal and minimal
entanglement of the final state, $C=\left\vert \sum_{j=1}^{4}b_{j}%
^{2}e^{2i\lambda_{j}}\right\vert $. We define a Lagrangian function%
\begin{equation}
L=\left\vert \sum_{j=1}^{4}b_{j}^{2}e^{2i\lambda_{j}}\right\vert ^{2}-\mu
_{1}\left(  \sum_{j=1}^{4}\left\vert b_{j}\right\vert ^{2}-1\right)  -\mu
_{2}\left(  \left\vert \sum_{j=1}^{4}b_{j}^{2}\right\vert ^{2}-C_{0}%
^{2}\right)  , \label{20}%
\end{equation}
where $\mu_{1}$ and $\mu_{2}$ are real. Differentiating gives%
\begin{equation}
2b_{j}e^{2i\lambda_{j}}\left(  \sum_{j=1}^{4}\left(  b_{j}^{\ast}\right)
^{2}e^{-2i\lambda_{j}}\right)  -\mu_{1}b_{j}^{\ast}-2\mu_{2}b_{j}\sum
_{j=1}^{4}\left(  b_{j}^{\ast}\right)  ^{2}=0. \label{21}%
\end{equation}
Multiplying $b_{j}$ and summing over $j$ gives%
\begin{equation}
\mu_{1}=2C^{2}-2\mu_{2}C_{0}^{2}\text{.} \label{22}%
\end{equation}
We write $\sum_{j=1}^{4}\left(  b_{j}^{\ast}\right)  ^{2}e^{-2i\lambda_{j}%
}=Ce^{2i\eta}$ and $\sum_{j=1}^{4}\left(  b_{j}^{\ast}\right)  ^{2}%
=C_{0}e^{2i\epsilon}$. Substituting them into Eq. (21), we get%
\begin{equation}
2b_{j}Ce^{2i\left(  \lambda_{j}+\eta\right)  }-\mu_{1}b_{j}^{\ast}-2\mu
_{2}b_{j}C_{0}e^{2i\epsilon}=0. \label{23}%
\end{equation}
One possible solution of the Eq. (23) is $b_{j}=0$. To find nonzero $b_{j}$,
we write $b_{j}=\beta_{j}e^{i\gamma_{j}}$. Then the Eq. (23) becomes%
\begin{equation}
C^{2}-\mu_{2}C_{0}^{2}-Ce^{2i\left(  \lambda_{j}+\eta+\gamma_{j}\right)  }%
+\mu_{2}C_{0}e^{2i\left(  \gamma_{j}+\epsilon\right)  }=0. \label{24}%
\end{equation}
If $\mu_{2}=0$, then $C^{2}-Ce^{2i\left(  \lambda_{j}+\eta+\gamma_{j}\right)
}=0$. Because $C$ is real, it will be $0$ or $1$. We have found the condition
that the entanglement of the final state is $0$ or $1$. So we assume that
$\mu_{2}$ is nonzero in the following. If there is only one nonzero
coefficient, the initial and final state will both be maximally entangled,
which is trivial. So there are at least two nonzero coefficients. Assume
$b_{j}$ and $b_{k}$ are nonzero. Similar to Eq. (24), we have%
\begin{equation}
C^{2}-\mu_{2}C_{0}^{2}-Ce^{2i\left(  \lambda_{k}+\eta+\gamma_{k}\right)  }%
+\mu_{2}C_{0}e^{2i\left(  \gamma_{k}+\epsilon\right)  }=0. \label{25}%
\end{equation}
Subtract Eq. (25) from (24), we get%
\begin{equation}
\left(  e^{2i\left(  \lambda_{j}+\eta+\gamma_{j}\right)  }-e^{2i\left(
\lambda_{k}+\eta+\gamma_{k}\right)  }\right)  C=\mu_{2}C_{0}\left(
e^{2i\left(  \gamma_{j}+\epsilon\right)  }-e^{2i\left(  \gamma_{k}%
+\epsilon\right)  }\right)  . \label{26}%
\end{equation}
Simplify Eq. (26), we get%
\begin{equation}
\sin\left(  \lambda_{j}-\lambda_{k}+\gamma_{j}-\gamma_{k}\right)
C=e^{i\left(  2\epsilon-2\eta-\lambda_{j}-\lambda_{k}\right)  }\mu_{2}%
C_{0}\sin\left(  \gamma_{j}-\gamma_{k}\right)  . \label{27}%
\end{equation}
We assume no $\lambda_{j}^{\prime}s$ are equal. Because $\mu_{2}C_{0}C$ is
nonzero and $\sin\left(  \gamma_{j}-\gamma_{k}\right)  $ can not be $0$, from
Eq. (27) we can get%
\begin{equation}
2\epsilon-2\eta-\lambda_{j}-\lambda_{k}=n\pi,\text{ }n\in Z. \label{28}%
\end{equation}
Because no $\lambda_{j}^{\prime}s$ are equal, there is only one pair of index
$\left(  j,k\right)  $ satisfying Eq. (28). So there are only one pair of
nonzero coefficients. Now Eq. (21) becomes%
\begin{equation}
2b_{j}e^{2i\lambda_{j}}\left(  \left(  b_{j}^{\ast}\right)  ^{2}%
e^{-2i\lambda_{j}}+\left(  b_{k}^{\ast}\right)  ^{2}e^{-2i\lambda_{k}}\right)
-\mu_{1}b_{j}^{\ast}-2\mu_{2}b_{j}\left(  \left(  b_{j}^{\ast}\right)
^{2}+\left(  b_{k}^{\ast}\right)  ^{2}\right)  =0. \label{29}%
\end{equation}
Substituting $b_{j}=\beta_{j}e^{i\gamma_{j}}$ and $b_{k}=\beta_{k}%
e^{i\gamma_{k}}$ into Eq. (29), we have%
\begin{equation}
2\beta_{j}^{2}+2\beta_{k}^{2}\cos\left(  \alpha+\beta\right)  -\mu_{1}%
-2\mu_{2}\left(  \beta_{j}^{2}+\beta_{k}^{2}\cos\alpha\right)  =0, \label{30}%
\end{equation}%
\begin{equation}
\sin\left(  \alpha+\beta\right)  =\mu_{2}\sin\alpha, \label{31}%
\end{equation}
where $\alpha=2\left(  \gamma_{j}-\gamma_{k}\right)  $, $\beta=2\left(
\lambda_{j}-\lambda_{k}\right)  $. Similarly we have%
\begin{equation}
2\beta_{k}^{2}+2\beta_{j}^{2}\cos\left(  \alpha+\beta\right)  -\mu_{1}%
-2\mu_{2}\left(  \beta_{k}^{2}+\beta_{j}^{2}\cos\alpha\right)  =0. \label{32}%
\end{equation}
Subtract Eq. (32) from (30), we get%
\begin{equation}
\left(  \beta_{j}^{2}-\beta_{k}^{2}\right)  \left[  \sin^{2}\left(
\frac{\alpha+\beta}{2}\right)  -\mu_{2}\sin^{2}\left(  \frac{\alpha}%
{2}\right)  \right]  =0. \label{33}%
\end{equation}
If $\beta_{j}^{2}\neq\beta_{k}^{2}$, then we have%
\begin{equation}
\left[  \sin^{2}\left(  \frac{\alpha+\beta}{2}\right)  -\mu_{2}\sin^{2}\left(
\frac{\alpha}{2}\right)  \right]  =0. \label{34}%
\end{equation}
From Eq. (31) and (34), we have $\beta/2=n\pi$, where $n$ is an integer. This
result contradicts with our assumption that no $\lambda_{j}^{\prime}s$ are
equal. So we have $\beta_{j}^{2}=\beta_{k}^{2}=1/2$. Now we rewrite the
concurrence of the initial state%
\begin{equation}
C_{0}^{2}=\left\vert \beta_{j}^{2}e^{2i\gamma_{j}}+\beta_{k}^{2}%
e^{2i\gamma_{k}}\right\vert ^{2}=\cos^{2}\left(  \gamma_{j}-\gamma_{k}\right)
. \label{35}%
\end{equation}
So%
\begin{equation}
\gamma_{j}-\gamma_{k}=n\pi\pm\arccos C_{0}\text{.} \label{36}%
\end{equation}
The concurrence of the final state%
\begin{align}
C^{2}  &  =\left\vert \beta_{j}^{2}e^{2i\left(  \gamma_{j}+\lambda_{j}\right)
}+\beta_{k}^{2}e^{2i\left(  \gamma_{k}+\lambda_{k}\right)  }\right\vert
^{2}\label{37}\\
&  =\cos^{2}\left(  \gamma_{j}-\gamma_{k}+\lambda_{j}-\lambda_{k}\right)
\nonumber\\
&  =\cos^{2}\left(  \arccos C_{0}\pm\left(  \lambda_{j}-\lambda_{k}\right)
\right)  .\nonumber
\end{align}
So the maximal possible concurrence of the final state is
\begin{equation}
C_{\max}=\max_{j,k}\left\vert \cos\left(  \arccos C_{0}+\left(  \lambda
_{j}-\lambda_{k}\right)  \right)  \right\vert . \label{38}%
\end{equation}
The minimal possible concurrence of the final state is%
\begin{equation}
C_{\min}=\min_{j,k}\left\vert \cos\left(  \arccos C_{0}+\left(  \lambda
_{j}-\lambda_{k}\right)  \right)  \right\vert . \label{39}%
\end{equation}
These results are derived when we assume no $\lambda_{k}^{\prime}s$ are equal.
This constraint can be removed by continuity and the reason is the same as
explained in Sec. III.

\section{conclusion}

In this paper we discuss the entanglement changing power of two-qubit unitary
operations without ancillas. A two-qubit unitary operation is charactered by
$\alpha_{1}$, $\alpha_{2}$, $\alpha_{3}$, or $\gamma_{1}$, $\gamma_{2}$,
$\gamma_{3},$ and $\gamma_{4}$. We consider a two-qubit unitary operation
along with arbitrary local unitary operations act on a two-qubit pure state,
whose entanglement is $C_{0}$. When the initial state is non-entangled
$\left(  C_{0}=0\right)  $, Kraus and Cirac \cite{kc} have calculated its
reachable maximal entanglement and we name it $C_{0\max}$. When the initial
state is maximally entangled, we calculate its reachable minimal entanglement
and we name it $C_{1\min}$. Then we give the conditions that the final state
can be maximally entangled and be non-entangled: when $C_{0}\geqslant
C_{1\min}$, the final state can be maximally entangled; when $C_{0}\leqslant
C_{0\max}$, the final state can be non-entangled. When the final state can not
be maximally entangled, we give its reachable maximal entanglement in Eq.
(38). When the final state can not be non-entangled, we give its reachable
minimal entanglement in Eq. (39). We think $C_{\max}$ and $C_{\min}$ represent
the entanglement changing power of two-qubit unitary operations. Now we write
our results in an unified form, which is much easier to operate.

If $\alpha_{1}+\alpha_{2}\geqslant\pi/4$ and $\alpha_{2}+\alpha_{3}%
\leqslant\pi/4$, $C_{\max}=1$ and $C_{\min}=0$.

If $\alpha_{1}+\alpha_{2}<\pi/4$ and $\alpha_{2}+\alpha_{3}\leqslant\pi/4$,
\[
C_{\max}=\cos\left(  \max\left[  \arccos C_{0}-2\left(  \alpha_{1}+\alpha
_{2}\right)  ,0\right]  \right)
\]
and%
\[
C_{\min}=\cos\left(  \min\left[  \arccos C_{0}+2\left(  \alpha_{1}+\alpha
_{2}\right)  ,\frac{\pi}{2}\right]  \right)  \text{.}%
\]

If $\alpha_{1}+\alpha_{2}\geqslant\pi/4$ and $\alpha_{2}+\alpha_{3}>\pi/4$,
\ \
\[
C_{\max}=\cos\left(  \max\left[  \arccos C_{0}-2\left(  \frac{\pi}{2}%
-\alpha_{2}-\alpha_{3}\right)  ,0\right]  \right)
\]
and%
\[
C_{\min}=\cos\left(  \min\left[  \arccos C_{0}+2\left(  \frac{\pi}{2}%
-\alpha_{2}-\alpha_{3}\right)  ,\frac{\pi}{2}\right]  \right)  \text{.}%
\]
It can be easily found that for the same initial state entanglement, the set
$\left[  C_{\min},C_{\max}\right]  $ of one two-qubit gate is a subset of
another's, or vice versa, and this relation will not change for arbitrary
$C_{0}$. So we can define an order of gates. We say two gates are equal if
their sets $\left[  C_{\min},C_{\max}\right]  $ are equal for any $C_{0}$. If
the set $\left[  C_{\min},C_{\max}\right]  $ of gate $U_{1}$\ is a true subset
of gate $U_{2}$'s for some $C_{0}$, we say $U_{1}<U_{2}$. Because swap gate
does not change entanglement when there is no ancilla, it is the smallest gate
according to this order.

\begin{center}
\textbf{Acknowledgement}
\end{center}

This work was funded by National Fundamental Research Program (Program No.
2001CB309300), National Natural Science Foundation of China.


\begin{thebibliography}{99}
\bibitem {explain1}Though Kraus and Cirac only calculated the maximal
entanglement of final states, it can easily find that the final states can be
non-entangled when initial states are product ones.
\end{thebibliography}
\end{document}